%% file: HgII.tex
\documentclass[twocolumn,aps,pra]{revtex4-1}

\usepackage{graphicx}
\usepackage{amsmath}
\usepackage{relsize}
\usepackage{dcolumn}

\include{phys_jdf}

\newcommand{\HgII}{Hg$^{+}$}

\newcolumntype{d}{D{.}{.}{-1}}   
\newcolumntype{b}{D{(}{\,(}{-1}} 

\newcommand{\E}[1]{\ensuremath{\times 10^{#1}}}
\newcommand{\cm}{\ensuremath{\textrm{cm}^{-1}}}

\newcommand{\bra}[1]{\ensuremath{\left< #1 \right|}}
\newcommand{\ket}[1]{\ensuremath{\left| #1 \right>}}

\newcommand{\vect}[1]{\boldsymbol{#1}}

\newcommand{\SigOne}{\ensuremath{\Sigma^{(1)}}}
\newcommand{\SigTwo}{\ensuremath{\Sigma^{(2)}}}
\newcommand{\SigOneTwo}{\ensuremath{\Sigma^{(1,2)}}}

\newcommand{\SigOneTwoThree}{\ensuremath{\Sigma^{(1,2,3)}}}
\newcommand{\SigVal}{\ensuremath{\Sigma_{val}}}
\newcommand{\SigValOne}{\ensuremath{\Sigma_{val}^{(1)}}}
\newcommand{\SigValTwo}{\ensuremath{\Sigma_{val}^{(2)}}}
\newcommand{\SigValThree}{\ensuremath{\Sigma_{val}^{(3)}}}
\newcommand{\SigValTwoThree}{\ensuremath{\Sigma_{val}^{(2,3)}}}

\newcommand{\eref}[1]{(\ref{#1})}

\newcommand{\Tref}[1]{Table~\ref{#1}}

\newcommand{\Sec}[1]{Section~\ref{#1}}

\begin{document}

\title{Particle-hole configuration interaction and many-body perturbation theory: application to Hg$^+$}

\author{J. C. Berengut}
\affiliation{School of Physics, University of New South Wales, Sydney, NSW 2052, Australia}

\date{25 May 2016}


\begin{abstract}
The combination of configuration interaction and many-body perturbation theory methods (CI+MBPT) is extended to non-perturbatively include configurations with electron holes below the designated Fermi level, allowing us to treat systems where holes play an important role. For example, the method can treat valence-hole systems like Ir$^{17+}$, particle-hole excitations in noble gases, and difficult transitions such as the $6s \rightarrow 5d^{-1}6s^2$ optical clock transition in Hg$^+$. We take the latter system as our test case for the method and obtain very good accuracy ($\sim 1\%$) for the low-lying transition energies. The $\alpha$-dependence of these transitions is calculated and used to reinterpret the existing best laboratory limits on the time-dependence of the fine-structure constant.
\end{abstract}

\maketitle

\section{Introduction}

The combination of configuration interaction and many-body perturbation theory (CI+MBPT) is a precise and flexible \emph{ab initio} method to calculate atomic properties of few-valence-electron atoms and ions~\cite{dzuba96pra}. It treats the valence-valence electron correlations using CI, while core-valence correlations are treated using MBPT by adding corrections to the radial integrals in CI. However several recent proposals have necessitated new methods of calculation in systems where holes play an important role and cannot be treated perturbatively. In this article we extend the CI+MBPT method to allow for arbitrary numbers of particles and holes while retaining the separation of correlation effects that allow us to apply MBPT. 

The CI+MBPT method was first developed to treat neutral thallium as a three-valence-electron atom, which gave an accuracy well below 1\% for the first few excitation energies~\cite{dzuba96pra}. Since then it has been remarkably successful in treating a wide variety of two and three-valence-electron atoms and ions, and even some with four (e.g.~\cite{berengut06pra,berengut13prl,ong13pra,savukov15pra}) and five valence electrons (Cr~II~\cite{berengut11pra}).
Generally, as the number of valence electrons increases the method becomes less effective and one must revert to usual CI and estimate the core-valence correlations some other way (see, e.g.~\cite{porsev07pra}). This is due to ever increasing `subtraction diagrams': one wishes to use ``spectroscopic'' orbitals calculated in the $V^N$ or $V^{N-1}$ approximation, but the MBPT expansion then contains large one-body diagrams representing the difference between the Dirac-Fock potentials used to calculate the orbital and that of the closed-shell core. A $V^{N-M}$ approximation was recommended in~\cite{dzuba05pra1} in order to simplify the MBPT calculation for the core-valence correlations, however then one must sacrifice the quality of the initial orbital which can be problematic particularly when treating open-shell systems.

The problem of large numbers of valence particles becomes even more acute when treating systems with nearly complete shells. It is here that the particle-hole CI+MBPT method presented in this paper can really help, since in this case one may take the Fermi level as being above the open shell and treating the valence holes of the atom or ion using CI. Such systems with holes include Ir$^{17+}$~\cite{windberger15prl}, proposed as an optical clock~\cite{berengut11prl}, as well as less exotic species such as Ni~II which is seen in quasar absorption spectra~\cite{murphy14mnras}. Currently these systems are treated as many-valence-electron systems using CI (e.g.~\cite{dzuba02praa}), but the methods presented in this work should allow for more accurate treatment.

The particle-hole CI+MBPT can also be used for calculating metastable states of noble gases. Previous works have calculated low-lying spectra of noble gases using this type of formalism~\cite{savukov02pra,savukov03jpb,savukov03jpb0,savukov12pra}, however the treatment presented here is more flexible in that it allows for additional particle-hole excitations and valence `spectators'.

This flexibility is a great strength of the particle-hole CI+MBPT method; we can take into account important excitations of electrons from below the Fermi level either using CI or MBPT, depending on how important the contribution of a shell is.  As an example, consider the original CI+MBPT system, neutral thallium~\cite{dzuba96pra}. In that work to get good accuracy the $6s^2$ electrons had to be included in CI due to their strong interaction with the valence electron. In effect, Tl was treated as a three-valence-electron system, while MBPT was used to get core-valence correlations with shells below $6s^2$. The cost was the inclusion of subtraction diagrams since the valence orbitals were calculated in the $V^{N-1}$ approximation while the $V^{N-3}$ core was frozen at CI level. With the current approach, we could keep the Fermi level above the $6s^2$ shell (i.e. using $V^{N-1}$) and still treat the excitations from the $6s^2$ shell non-perturbatively using particle-hole CI.

In this paper we test our method by calculating low-lying transitions in the \HgII\ ion, important because laser-cooled \HgII\ ions are used for both microwave~\cite{berkeland98prl} and optical~\cite{diddams01sci} frequency standards. Calculations of energy levels, blackbody radiation shifts and oscillator strengths were previously performed using both third-order relativistic many-body perturbation theory and the single-double all-order method~\cite{simmons11pra}, but crucially the optical clock transition was not accessible using these methods. To lowest order the $6s \rightarrow 5d^{-1}6s^2$ clock transition can be described as a particle-hole excitation, with the valence $6s$ electron a spectator. It is precisely this sort of system that our method is designed to treat.

One important use of the \HgII\ optical clock is to constrain potential drift in the value of the fine-structure constant, $\alpha = e^2/\hbar c$. Measurements of the frequency ratio of the $^{199}$Hg$^+$ and Al$^+$ optical atomic clocks were taken several times over the course of a year~\cite{rosenband08sci}. The \HgII\ clock frequency is highly sensitive to the value of $\alpha$, while the Al$^+$ is relatively insensitive. The resulting limit on $\dot\alpha/\alpha$ remains the tightest laboratory constraint on variations of fundamental constants, but calculations of the $\alpha$-dependence of the \HgII\ transition are based only on configuration interaction calculations treating the ion as an 11-valence-electron system~\cite{dzuba99pra,dzuba08pra0}. In this work we use the particle-hole CI+MBPT method to calculate the transition frequencies and $\alpha$-dependence of the low-lying transitions in \HgII, including the clock transition. We use this to reinterpret the measurements of the \HgII/Al$^+$ frequency ratio to obtain updated laboratory limits on $\dot\alpha/\alpha$.

This work is organised as follows. In \Sec{sec:CI} we introduce the particle-hole CI formalism and compare it against the usual `electron-only' CI for our \HgII\ test case. As expected, both methods give the same transition energies. We then add core-valence correlations using MBPT in \Sec{sec:CI+MBPT}, which shows that only in the particle-hole formalism does the addition of MBPT improve the results for \HgII. In \Sec{sec:val} we add some additional MBPT diagrams representing valence-valence correlations that arise in the particle-hole formalism. Finally in \Sec{sec:alpha} we calculate the $\alpha$-dependence of the \HgII\ transitions. Atomic units ($\hbar = m_e = |e| = 1$) are used throughout.

\section{Configuration interaction with holes}
\label{sec:CI}

To start our calculation, we solve the self-consistent Dirac-Fock equations for the core electrons,
\begin{equation}
\hat h^\textrm{DF} \ket{m} = \varepsilon_m \ket{m}
\end{equation}
where
\begin{equation}
\hat h^\textrm{DF} = c \vect{\alpha}\cdot\vect{p} + (\beta - 1) c^2 - V^{N_\textrm{core}} (r) .
\end{equation}
The potential $V^{N_\textrm{core}}$ includes the nuclear potential ($Z/r$ outside the nucleus and with finite-size corrections within it) and the electronic potential with both direct and exchange parts of the core electrons included in the self-consistent Hartree-Fock procedure. For the present Hartree-Fock calculation we include 78 core electrons in the configuration [Xe]\,$4f^{14}\,5d^{10}$.
Here all shells are closed, but in general we can sometimes obtain better starting orbitals by including a partially-filled closed shell as was done in previous works, e.g.~\cite{berengut08jpb,berengut11pra}. However, we must then include MBPT subtraction diagrams (see \Sec{sec:CI+MBPT}).

We then generate a single-particle basis set \ket{i} by diagonalising a set of B-splines over $\hat h^\textrm{DF}$~\cite{johnson88pra}. The resulting orbitals include core and valence orbitals and a large number of virtual orbitals (pseudostates), which we reduce in number by excluding those with the highest energy. 

The many-electron basis is formed from configuration state functions (CSFs) denoted below with capital letters \ket{I}. Slater determinants are first formed from the orbitals \ket{i}. All Slater determinants with fixed angular momentum projection $M$ corresponding to a configuration are diagonalised over the $\hat J^2$ operator, giving us CSFs with fixed angular momentum $J$ and projection $M$.

The many-electron Hilbert space is separated into sub-spaces $\mathcal{P}$ and its complement $\mathcal{Q}$ ($\mathcal{P} + \mathcal{Q} = 1$). CSFs in the $\mathcal{P}$ space are included in the configuration interaction procedure directly, while those in the $\mathcal{Q}$ space are treated using many-body perturbation theory.
In the CI method the many-electron wavefunction $\psi$ is expressed as a linear combination of CSFs from the subspace $\mathcal{P}$ only:
\begin{equation}
\psi = \sum_{I \in \mathcal{P}} C_I \ket{I} .
\end{equation}
The coefficients $C_I$ are obtained from the matrix eigenvalue problem
\begin{equation}
\sum_{J \in \mathcal{P}} H_{IJ} C_J = E C_I
\end{equation}
where $H_{IJ}$ is the matrix element of the exact Dirac-Coulomb Hamiltonian operator $\mathcal{H}$ projected onto the model subspace using the projection operator $\hat{\mathcal{P}}$:
\begin{align}
\hat{\mathcal{P}}\mathcal{H}\hat{\mathcal{P}}
	= & \sum_i  c\, \vect{\alpha}\cdot\vect{p}_i + (\beta - 1) c^2 + e_i V^{N_\textrm{core}}(r_i) \nonumber \\
	& + \sum_{i < j} \frac{e_i e_j}{|\vect{r}_i - \vect{r}_j|} .
\label{eq:PHP}
\end{align}
Here $i$ and $j$ run over the valence electrons and holes, and $e_i$ is $-1$ if $i$ is an electron state (above the Fermi level) and $+1$ if it is a hole. The resulting energies $E$ are therefore calculated with respect to the Fermi level; that is, the closed shell core has $E = 0$.

We introduce a second quantization notation to separate $\mathcal{H}$ into one and two-body operators (see~\cite{berengut06pra} for details)
\begin{gather}
\label{eq:H1}
\mathcal{H}^{(1)} = \sum_{ij} \{ a_i^\dagger a_j \} \bra{i} \hat h^\textrm{CI} \ket{j} \\
\label{eq:H2}
\mathcal{H}^{(2)} = \frac{1}{2} \sum_{ijkl} \{ a_i^\dagger a_j^\dagger a_l a_k \} \bra{ij} r_{12}^{-1} \ket{kl} \,.
\end{gather}
Here $a_i^\dagger$ and $a_i$ are electron creation and annihilation operators, and the brackets $\{...\}$ denote normal ordering with respect to the closed-shell core.

In previous works all CSFs in the valence space $\mathcal{P}$ had the same number of valence electrons. Our code, however, allows for additional particle-hole pairs, provided that the total fermion number is conserved. For example, our calculations of \HgII\ include CSFs based on configurations $\ket{6s}$, $\ket{5d^{-1} 6s^2}$ and $\ket{5d^{-2} 6s\,6p^2}$ (among many others). We express these using second quantisation with respect to the Fermi level; the Wick contractions required to calculate matrix elements $H_{IJ}$ were implemented in our atomic code AMBiT~\cite{berengut08jpb}.

To test our code, we compare our particle-hole CI calculation for \HgII\ with a traditional CI calculation.
To form the set of $\mathcal{P}$-space configurations used we start with the leading configurations $\ket{6s}$, $\ket{6p}$, $\ket{5d^{-1} 6s^2}$, $\ket{5d^{-1} 6p^2}$ and $\ket{5d^{-1} 6s\,6p}$. From these we take single electron excitations up to $16spdf$ and allow an additional hole excitation in the $5d$ shell only. (The notation $16spdf$ refers to the highest principal quantum number for each wave, in this case \mbox{1\,--\,$16s$}, \mbox{2\,--\,$16p$}, etc. Note that higher orbitals are pseudostates.)
We then allow a second electron excitation up to $10spdf$. For the traditional CI calculation, where the $5d^{10}$ shell is taken as valence above the Fermi level, this is equivalent to allowing single excitations from the leading configurations up to $16spdf$ and double excitations up to $10spdf$, but ensuring at least 8 electrons remain in the $5d$ shell. The resulting CI matrices are rather large; for example, the $J = 5/2$ odd-parity matrix includes 191\,511 CSFs. However, the CI configuration set ($\mathcal{P}$ space) cannot be said to be saturated even in this case.

We present the comparison in \Tref{tab:CI}. The particle-hole method returns the single-electron binding energy of each low-lying level. The electron-only CI method returns the binding energy for 11 electrons (i.e. back to the Fermi level below $5d^{10}$). Both methods should give exactly the same level spacings; in fact, they are slightly different due to small numerical errors in the integration routines. Thus it is here that we see the first advantage of the particle-hole CI method: it is numerically stable because it doesn't rely on large cancelation of binding energies.

\begin{table}
\caption{\label{tab:CI} Configuration interaction calculations of \HgII\ using traditional CI (electrons only) and the particle-hole CI. $E$ is the valence binding energy and $\Delta$ is the excitation energy relative to the $5d^{10}\,6s$ ground state. All energies in \cm.}
\begin{ruledtabular}
\begin{tabular}{lrrrrr}
Level & J & \multicolumn{2}{c}{Electrons only} & \multicolumn{2}{c}{Electrons \& Holes} \\
 & & \multicolumn{1}{c}{$E$} & \multicolumn{1}{c}{$\Delta$} & \multicolumn{1}{c}{$E$} & \multicolumn{1}{c}{$\Delta$} \\
\hline
$6s\ ^2$S
 & 1/2 & $-9085835$ & 0 & $-149653$ & 0 \\
$5d^{-1}\,6s^2\ ^2$D
 & 5/2 & $-9047981$ & 37854 & $-111840$ & 37814 \\
 & 3/2 & $-9032845$ & 52990 & $-96704$ & 52949 \\
$6p\ ^2$P$^o$
 & 1/2 & $-9036118$ & 49717 & $-100187$ & 49466 \\
 & 3/2 & $-9027966$ & 57869 & $-92112$ & 57541 \\
$5d^{-1}\,6s\,6p$
 & 5/2 & $-9006616$ & 79219 & $-70704$ & 78949 \\
$5d^{-1}\,6s\,6p$
 & 7/2 & $-9001801$ & 84034 & $-65907$ & 83746 \\
 & 5/2 & $-9001291$ & 84544 & $-65396$ & 84257 \\
 & 3/2 & $-8999704$ & 86131 & $-63846$ & 85807 \\
\end{tabular}
\end{ruledtabular}
\end{table}

\section{CI + MBPT}
\label{sec:CI+MBPT}

Our implementation of the CI+MBPT method~\cite{dzuba96pra} is described in detail elsewhere~\cite{berengut06pra}. Omitting mathematical details, we write the exact Hamiltonian $\mathcal{H}$ in the subspace $\mathcal{P}$ using the Feshbach operator, which yields the exact energy when operating on the model function $\Psi_P = \hat{\mathcal{P}}\Psi$:
\begin{equation}
\left( \hat{\mathcal{P}}\mathcal{H}\hat{\mathcal{P}} + \Sigma(E) \right) \Psi_P = E \Psi_P.
\end{equation}
We can then generate a perturbation expansion for $\Sigma$ in the residual Coulomb interaction, which to second order can be written in matrix form as
\begin{equation}
\label{eq:sigma}
\Sigma_{IJ} = \sum_{M\in\mathcal{Q}} \frac{\bra{I}H\ket{M}\bra{M}H\ket{J}}{E - E_M}
\end{equation}
where $I$ and $J$ enumerate CSFs from the model subspace $\mathcal{P}$. The final equation of the CI+MBPT method can be expressed as
\begin{equation}
\label{eq:CI+MBPT}
\sum_{J \in \mathcal{P}} \left(H_{IJ} + \sum_{M\in\mathcal{Q}} \frac{\bra{I}H\ket{M}\bra{M}H\ket{J}}{E - E_M} \right)C_J = E C_I \,.
\end{equation}
Thus the method includes correlations with configurations in the $\mathcal{Q}$ space by changing the matrix elements in the $\mathcal{P}$-space CI calculation.
In practice, we simplify this procedure by modifying the one and two-particle radial integrals in Eqs.~\eref{eq:H1} and~\eref{eq:H2}.
A diagrammatic technique for calculating $\Sigma$ is presented in \cite{berengut06pra} along with explicit expressions for the radial integrals.

In \Tref{tab:CI+MBPT} we compare CI+MBPT calculations using the traditional CI method and the particle-hole method. In both CI calculations we consider orbitals below $5d^{10}$ as frozen (i.e. there are no configurations with holes in the $5s^2$, $5p^6$, $4d^{10}$ and $4f^{14}$ shells, or those below them). Correlations with the frozen core are therefore treated using MBPT; excited orbitals up to $30spdfgh$ are included in the MBPT diagrams. \SigOne\ calculations include MBPT modifications to the one-body integrals of \eref{eq:H1}; \SigOneTwo\ includes MBPT in both one-body and the two-body integrals \eref{eq:H2}; while \SigOneTwoThree\ also includes effective three-body core-valence integrals that occur in second order of MBPT (see~\cite{berengut08jpb} for details).

\begin{table*}
\caption{\label{tab:CI+MBPT} CI+MBPT calculations of excitation energies for \HgII\ using traditional CI (electrons only) and the particle-hole CI. Calculations including MBPT in one-body and two-body integrals are labelled \SigOne\ and \SigOneTwo\ respectively, while \SigOneTwoThree\ includes effective three-body interactions. All energies in \cm.}
\begin{ruledtabular}
\begin{tabular}{lrrrrrrrrr}
Level & J & CI & \multicolumn{3}{c}{Electrons only} & \multicolumn{3}{c}{Electrons \& Holes} & Expt. \\
 & & & \SigOne & \SigOneTwo & \SigOneTwoThree & \SigOne & \SigOneTwo & \SigOneTwoThree \\
\hline
$6s\ ^2$S
 & 1/2 & 0 & 0 & 0 & 0 & 0 & 0 & 0 & 0\\
$5d^{-1}\,6s^2\ ^2$D
 & 5/2 & 37814 & 17957 & 28432 & 32205 & 27197 & 34683 & 34721 & 35515 \\
 & 3/2 & 52949 & 30698 & 44362 & 48001 & 41736 & 50095 & 50027 & 50556 \\
$6p\ ^2$P$^o$
 & 1/2 & 49466 & 49328 & 49356 & 51137 & 53494 & 52010 & 51908 & 51486 \\
 & 3/2 & 57541 & 59358 & 55272 & 59952 & 62948 & 61297 & 61188 & 60608 \\
$5d^{-1}\,6s\,6p$
 & 5/2 & 78949 & 60024 & 69157 & 74890 & 71945 & 78727 & 78975 & 79705 \\
$5d^{-1}\,6s\,6p$
 & 7/2 & 83746 & 64567 & 74422 & 79825 & 77009 & 83164 & 83206 & 84212 \\
 & 5/2 & 84257 & 65021 & 75326 & 80413 & 77442 & 83606 & 83727 & 84836 \\
 & 3/2 & 85807 & 66882 & 76143 & 81850 & 79169 & 85142 & 85136 & 86178 \\
\end{tabular}
\end{ruledtabular}
\end{table*}

Unlike in the pure CI calculations presented in \Tref{tab:CI}, there is no reason in this case that the two calculations should give the same result. Indeed, one of the purposes of this work is to avoid the large subtraction diagrams in \SigOne\ that are partially cancelled by terms in \SigTwo\ (see~\cite{berengut11pra} for details). Subtraction diagrams are not present in the particle-hole calculation since in that case $\hat h^\textrm{CI} = \hat h^\textrm{DF}$ and there are no off-diagonal matrix elements of \eref{eq:H1} (at least until MBPT corrections are included). \Tref{tab:CI+MBPT} shows that the accuracy of calculation of low-lying levels is improved by MBPT in the particle-hole calculation, but not in the traditional electron-only calculation.

\section{MBPT corrections to valence-valence integrals}
\label{sec:val}

In previous implementations of CI+MBPT, the $\mathcal{Q}$ space is defined to include all configurations with holes in the core. Since the valence space doesn't include holes, this was a clear delineation. Now that we can include configurations with holes in the CI calculation, we must redefine the $\mathcal{Q}$ space. In this work we take the $\mathcal{Q}$ space to include any configurations with holes below the $5d^{10}$ shell (not including it) or with electron excitations above the valence space.

\begin{figure}[tb]
\caption{\label{fig:VV1} One-body valence-valence subtraction diagram \SigValOne.}
\includegraphics{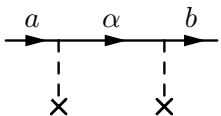}
\end{figure}
\begin{figure}[htb]
\caption{\label{fig:VV2} Two-body valence-valence diagrams \SigValTwo. The subtraction diagram (b) represents four diagrams, with the complementary diagrams obtained by reflection in the horizontal and vertical planes.}
\includegraphics{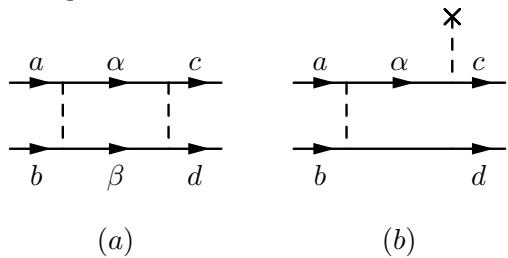}
\end{figure}
\begin{figure}[htb]
\caption{\label{fig:VV3} Effective three-body valence-valence diagram \SigValThree.}
\includegraphics{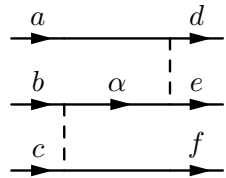}
\end{figure}

The particle-hole CI+MBPT method then allows for an additional type of diagram that has no additional core holes, but does have electron excitations outside the valence space. At second order in the residual Coulomb interaction these valence-valence diagrams occur in the one-body, two-body, and effective three-body operators, as shown in Figs.~\ref{fig:VV1} -- \ref{fig:VV3}. In these diagrams the external lines marked $a$, $b$, ... are valence electrons or holes, while the internal lines marked $\alpha$, $\beta$ are virtual electron orbitals outside the CI valence space. Diagrams with external field lines (Figs.~\ref{fig:VV1} and \ref{fig:VV2}(b)) are known as subtraction diagrams since the one-body external field operator is $\hat h^\textrm{CI} - \hat h^\textrm{DF}$. In the current work these diagrams are zero since $\hat h^\textrm{CI} = \hat h^\textrm{DF}$.
Note that hole-hole diagrams, where the virtual electron orbitals ($\alpha$, $\beta$) are replaced with non-valence holes, are included already; e.g. Fig.~3(f) in~\cite{berengut06pra} is Fig.~\ref{fig:VV2}(a) with $\alpha$ and $\beta$ replaced with hole states.

Including valence-valence diagrams allows us to significantly reduce the size of the CI calculation. \Tref{tab:VV} shows the results of our smaller CI calculation in which, from the same set of leading configurations used in Sections~\ref{sec:CI} and~\ref{sec:CI+MBPT}, we allow double electron excitations up to $10spdf$ and single hole excitations in $5d$ only (as before). In this case the matrix size for the $J = 5/2$ odd-parity calculation is 57\,879 --- much smaller than in the previous calculations which included additional single-electron excitations up to $16spdf$. 

\begin{table}[htb]
\caption{\label{tab:VV} Particle-hole CI+MBPT calculations of \HgII\ using a smaller basis for CI ($10spdf$) and \SigOneTwoThree\ (third column). The effect of adding \SigVal\ is shown in the fourth and fifth columns, which add \SigValTwo\ (Fig. 2(a)) and \SigValTwoThree\ (Figs. 2(a) and 3), respectively. All energies in \cm. }
\begin{ruledtabular}
\begin{tabular}{lrrrrr}
Level & J & \multicolumn{1}{c}{\SigOneTwoThree} &  \multicolumn{1}{c}{$+\SigValTwo$} &  \multicolumn{1}{c}{$+\SigValTwoThree$} & Expt. \\
\hline
$6s\ ^2$S
 & 1/2 & 0 & 0 & 0 & 0\\
$5d^{-1}\,6s^2\ ^2$D
 & 5/2 & 32418 & 31940 & 35121 & 35515 \\
 & 3/2 & 47840 & 47201 & 50446 & 50556 \\
$6p\ ^2$P$^o$
 & 1/2 & 53279 & 53654 & 53693 & 51486 \\
 & 3/2 & 62671 & 63093 & 63188 & 60608 \\
$5d^{-1}\,6s\,6p$
 & 5/2 & 79221 & 80806 & 85917 & 79705 \\
$5d^{-1}\,6s\,6p$
 & 7/2 & 83211 & 83742 & 88486 & 84212 \\
 & 5/2 & 83877 & 84534 & 89031 & 84836 \\
 & 3/2 & 85018 & 85626 & 90558 & 86178 \\
\end{tabular}
\end{ruledtabular}
\end{table}

\Tref{tab:VV} shows that including valence-valence diagrams can bring smaller CI+MBPT calculations more in line with the larger ones, although clearly this can overshoot the experimental values. This may point to the possibility that a `converged' CI calculation using second-order MBPT with no valence-valence diagrams might be similarly discrepant with the experiment. In any case the results strongly suggest that valence-valence diagrams can be of help in cases where the CI matrix grows very rapidly and it is not possible to even approach convergence.

\section{Dependence on the fine-structure constant}
\label{sec:alpha}

We have calculated the dependence of the levels on the fine-structure constant $\alpha$, usually expressed with the $q$ value defined by
\begin{equation}
\omega(\alpha) = \omega_0 + q x,
\end{equation}
where $x = (\alpha/\alpha_0)^2 -1$, and $\omega_0$ is the laboratory energy with $\alpha$ given by its present-day value $\alpha_0$.
To calculate $q$ we vary $\alpha$ directly in the code and extract the numerical derivative over $x$.

Our results are presented in \Tref{tab:alpha}. The value quoted for this work is the average of two methods: the large CI+\SigOneTwoThree\ from \Sec{sec:CI+MBPT} and the CI+\SigOneTwoThree+\SigValTwoThree\ calculation of \Sec{sec:val}; both calculations give energies that are close to experiment and $q$ values that are highly consistent. The error quoted is roughly half the difference between the two calculations: these should be taken as indicative errors only.

\begin{table}[tb]
\caption{\label{tab:alpha} Calculated dependence on the fine-structure constant, $q$ (\cm). }
\begin{ruledtabular}
\begin{tabular}{lrrbrr}
Level & J & \multicolumn{1}{c}{$E$ (\cm)} & \multicolumn{2}{c}{$q$ (\cm)} \\
 & & Expt. & \multicolumn{1}{c}{This work} & \multicolumn{1}{c}{Other}  \\
\hline
$6s\ ^2$S
 & 1/2 & 0 & 0 \\
$5d^{-1}\,6s^2\ ^2$D
 & 5/2 & 35515 & -50667(600) & $-56670$\footnotemark[1] \\
 &       &           &                       & $-52200$\footnotemark[2] \\
 & 3/2 & 50556 & -35960(600) & $-44000$\footnotemark[1] \\
 &       &           &                       & $-37700$\footnotemark[2] \\
$6p\ ^2$P$^o$
 & 1/2 & 51486 & 15907(600)\\
 & 3/2 & 60608 & 28958(900)\\
$5d^{-1}\,6s\,6p$
 & 5/2 & 79705 & -35788(400)\\
$5d^{-1}\,6s\,6p$
 & 7/2 & 84212 & -34233(400)\\
 & 5/2 & 84836 & -33158(300)\\
 & 3/2 & 86178 & -32654(700)\\
\end{tabular}
\end{ruledtabular}
\footnotetext[1]{Ref.~\cite{dzuba99pra}}
\footnotetext[2]{Ref.~\cite{dzuba08pra0}}
\end{table}

Of particular interest is the $6s\ ^2\textrm{S}_{1/2}$ -- $5d^{-1}\,6s^2\ ^2\textrm{D}_{5/2}$ transition at $\omega = 35515\,\cm$, which is the reference transition for the NIST \HgII\ clock~\cite{diddams01sci}. This transition was compared with the 37393\,\cm Al$^+$ clock~\cite{schmidt05sci,rosenband07prl} several times over the course of a year, and the frequency ratio $\nu_{\textrm{Al}^+}/\nu_{\textrm{Hg}^+}$ was found to vary by $(-5.3\pm7.9)\E{-17}/\textrm{year}$~\cite{rosenband08sci}. With the $q$ value given in \Tref{tab:alpha} for the \HgII\ clock transition, and taking the Al${^+}$ $q$ value from Ref.~\cite{angstmann04pra0}, we find that the sensitivity of the ratio to a fractional change in $\alpha$ is $-2.861\,(34)$. Therefore we extract an updated limit on time-variation of $\alpha$ of $\dot\alpha/\alpha = (-1.8\pm2.8)\E{-17}/\textrm{year}$.

\section{Conclusion}

We have presented a particle-hole CI+MBPT theory that provides more flexibility than previous versions. In particular, it should be able to accurately calculate systems that are better treated with holes; access particle-hole excitations; and give us the choice to treat correlations with filled core shells either perturbatively using MBPT or non-perturbatively using CI. We have applied the method to low-lying transition energies in \HgII, including the optical clock transition which to lowest-order is a particle-hole excitation.

The current limitation of our method is in the energy denominators of Eq.~\eref{eq:sigma}. In keeping with our previous CI+MBPT methods, we have employed Brillouin-Wigner perturbation theory (see~\cite{berengut06pra} for details), but this cannot be an accurate treatment for all levels. Methods to treat the energy denominators to higher order have been developed, such as including $\partial\Sigma/\partial E$~\cite{dzuba96pra} or simple addition of an offset in the denominator~\cite{kozlov99os,berengut08jpb}, and these may improve our accuracy in the future. The particle-hole CI formalism can also be combined with other methods used to calculate core-valence correlations, such as the all-order correlation potential~\cite{dzuba89pla,dzuba08pra1} or singles-doubles coupled-cluster~\cite{safronova09pra} methods.

\section*{Acknowledgements}

I thank J. S. M. Ginges for her careful reading of this paper, and
V. A. Dzuba, V. V. Flambaum, C. Harabati, I. D. Leroux, and P. O. Schmidt for useful discussions.
This work was supported in part by the Australian Research Council, grant DE120100399.

\bibliography{references}

\end{document}

%% file: phys_jdf.tex
\def\jpb{J. Phys. B}

\def\mnras{Mon. Not. R. Astron. Soc.}

\def\os{Opt. Spectrosc.}

\def\pla{Phys. Lett. A}

\def\pra{Phys. Rev. A}

\def\prl{Phys. Rev. Lett.}

\def\sci{Science}